\documentclass[11pt,a4paper]{article}
\usepackage[margin=1in]{geometry}
\usepackage{amsmath,amssymb,amsthm}
\usepackage{graphicx}
\usepackage{natbib}
\usepackage{hyperref}
\usepackage{url,xcolor}
\usepackage{booktabs} 

\hypersetup{
    colorlinks=true,
    linkcolor=blue,
    filecolor=magenta,      
    urlcolor=cyan,
    citecolor=blue,
}

\newtheorem{theorem}{Theorem}[section]

\newtheorem{proposition}[theorem]{Proposition}
\theoremstyle{definition}
\newtheorem{definition}{Definition}[section]
\theoremstyle{remark}
\newtheorem{remark}{Remark}[section]

\newcommand{\R}{\mathbb{R}}

\newcommand{\norm}[1]{\left\lVert#1\right\rVert}

\newcommand\eg{\textit{e.g.,}}

\title{Topolow: Force-Directed Euclidean Embedding of Dissimilarity Data with Robustness Against Non-Metricity and Sparsity}
\author{
  Omid Arhami*\\
  \texttt{omid.arhami@uga.edu}
  \and
  Pejman Rohani\\
  \texttt{rohani@uga.edu}
}
\date{August, 2025}

\begin{document}

\maketitle

\begin{abstract}
The problem of embedding a set of objects into a low-dimensional Euclidean space based on a matrix of pairwise dissimilarities is fundamental in data analysis, machine learning, and statistics. However, the assumptions of many standard analytical methods are violated when the input dissimilarities fail to satisfy metric or Euclidean axioms. We present the mathematical and statistical foundations of Topolow, a physics-inspired, gradient-free optimization framework for such embedding problems. Topolow is conceptually related to force-directed graph drawing algorithms but is fundamentally distinguished by its goal of quantitative metric reconstruction. It models objects as particles in a physical system, and its novel optimization scheme proceeds through sequential, stochastic pairwise interactions, which circumvents the need to compute a global gradient and provides robustness against convergence to local optima, especially for sparse data. Topolow maximizes the likelihood under a Laplace error model, robust to outliers and heterogeneous errors, and properly handles censored data. Crucially, Topolow does not require the input dissimilarities to be metric, making it a robust solution for embedding non-metric measurements into a valid Euclidean space, thereby enabling the use of standard analytical tools. We demonstrate the superior performance of Topolow compared to standard Multidimensional Scaling (MDS) methods in reconstructing the geometry of sparse and non-metric data. This paper formalizes the algorithm, first introduced as Topolow in the context of antigenic mapping in \citet{arhami2025topolow} (open access), with emphasis on its Euclidean embedding and mathematical properties for a broader audience. The general-purpose function \texttt{Euclidify} is available in the R package \texttt{topolow}.
\end{abstract}

\section{Introduction}

The task of representing the similarity or dissimilarity of objects in a low-dimensional Euclidean space is a cornerstone of data analysis. These objects can range from abstract concepts, such as customer perception of different brands \citep{cooper1983review, borg2018applied}, to complex biological measurements such as protein folding \citep{holm1996mapping}, or information constructs that do not necessarily satisfy the metric axioms and, hence, are incompatible with many standard statistical or machine learning tools \citep{duin2005dissimilarity}. When dissimilarity data violate metric axioms, they cannot be embedded in a vector space where linear algebraic operations are valid. The fundamental issue is that coordinate-based methods assume data points exist in a vector space equipped with a valid inner product structure. This inner product ⟨·,·⟩ --used in many models, from linear regression to artificial intelligence systems-- induces a metric satisfying all axioms \citep{burago2001course}.

Given a set of $n$ objects $S = \{s_1, \dots, s_n\}$ and an $n \times n$ matrix $D$ of pairwise dissimilarities $D_{ij}$, the goal of Euclidean embedding is to find a set of coordinates $X = \{x_1, \dots, x_n\}$, where $x_i \in \R^N$ for some dimension $N$, such that the Euclidean distance $d(x_i, x_j) = \norm{x_i - x_j}$ is a faithful representation of $D_{ij}$.

A standard method for this task is Multidimensional Scaling (MDS). Classical MDS provides a deterministic, analytical solution via eigenvalue decomposition \citep{torgerson1952multidimensional, gower1966some}, while iterative MDS variants use numerical optimization to minimize a "stress" function, typically defined as the sum of squared differences between observed dissimilarities and the embedded distances \citep{kruskal1964multidimensional}:
\begin{equation}
\text{Stress}(X) = \sum_{i<j, (i,j) \in \mathcal{O}} \left( D_{ij} - \norm{x_i - x_j} \right)^2,
\label{eq:stress}
\end{equation}
where $\mathcal{O}$ is the set of pairs with observed dissimilarities. These gradient-based methods face significant challenges when the dissimilarity matrix is highly sparse or when the dissimilarities themselves do not satisfy metric axioms, leading to unstable solutions or convergence to poor local minima \citep{bravo2002}.

\subsection{The Challenge of non-metric Dissimilarities}
The fundamental challenge of confronting and resolving non-metric structures has deep roots in the history of geometry. The logical foundations of non-Euclidean space were first systematically investigated by the 11th-century Persian mathematician Omar Khayyam in his treatise "\textit{The Difficulties in the Postulates of Euclid's Elements}" \citep{khayyam1936difficulties, struik1958omar}. Many standard data analysis techniques, particularly those from linear algebra, are fundamentally incompatible with non-metric data.

The most prominent example is Principal Component Analysis (PCA). PCA operates on the variance-covariance structure of data, which is derived from the inner products of vectors representing the data points. The Torgerson-Gower theorem \citep{torgerson1952multidimensional, gower1966some} provides the connection between a dissimilarity matrix $D$ and the required inner product (Gram) matrix $B$: $B = -\frac{1}{2} J D^2 J$, where $J$ is a centering matrix. This transformation is only valid if $D$ is Euclidean, i.e., the points it describes can be perfectly embedded in a Euclidean space of some dimension. If it is not, the resulting Gram matrix $B$ will not be positive semidefinite (PSD), meaning it will possess one or more negative eigenvalues \citep{gower1985}. Since eigenvalues in PCA correspond to the variance along each principal component, a negative eigenvalue implies a dimension with "imaginary" variance—a geometric and statistical absurdity that invalidates the analysis.

Similarly, clustering algorithms are affected by non-metric dissimilarities. Centroid-based methods, such as $k$-means, become unusable as they require coordinates to compute means. Distance-based methods like hierarchical clustering will run, but can produce counter-intuitive results if the data violate the triangle inequality ($D_{ik} > D_{ij} + D_{jk}$), a key metric property. Such violations disrupt the fundamental notion of "closeness" upon which these algorithms depend.

Therefore, for a vast significant range of common data, a critical preliminary step is to find a faithful Euclidean representation. An embedding algorithm that can robustly map non-metric dissimilarities into a valid Euclidean space acts as an essential bridge, making the data amenable to the entire suite of standard analytical tools. Topolow is designed to be this bridge.\newline

To address these challenges, we discuss the mathematical framework of Topolow, an algorithm first introduced by \citet{arhami2025topolow} for application in antigenic cartography and the study of viral evolution. Topolow adapts the physical analogy of force-directed methods \citep{fruchterman1991, kamada1989, kobourov2012} but re-purposes it for the distinct goal of high-fidelity metric reconstruction from potentially sparse, noisy, biased, and non-metric dissimilarity data. We demonstrate that:
\begin{enumerate}
    \item The algorithm operates on a general dissimilarity space, requiring only non-negative dissimilarities, without assuming symmetry or the triangle inequality.
    \item The optimization procedure is gradient-free and robust to local optima.
    \item The choice of a Mean Absolute Error (MAE) loss function is statistically justified, corresponding to a maximum likelihood estimate under a Laplace error model, which is well-suited for noisy data with outliers and censored measurements.
    \item The framework includes a principled, likelihood-based method for selecting the optimal embedding dimension $N$.
\end{enumerate}
We thus position Topolow as a powerful, general-purpose mathematical tool for Euclidean embedding, dimensionality reduction, and data visualization.

\section{The Topolow Framework}

\subsection{Problem Definition}
Let $P = \{p_1, \dots, p_m\}$ be a set of $m$ objects. Let $D$ be an $m \times m$ matrix where entries $D_{ij}$ represent a measure of dissimilarity between objects $i$ and $j$. The problem is to find a configuration of points $X = \{\textbf{x}_1, \dots, \textbf{x}_{m}\}$, where each $\textbf{x}_k \in \R^N$, that best represents the relationships in $D$. The dimension $N$ is also a parameter to be determined.

\subsection{From Similarity to Dissimilarity}
Topolow operates on dissimilarities (distances). If raw data measures similarity, where higher values imply greater affinity, a transformation is used to convert a similarity measure $S_{ij}$ into a dissimilarity $D_{ij}$.

\begin{definition}[Similarity-to-Dissimilarity Transformation]
Let $S_{ij}$ be the similarity of object $i$ to object $j$. The corresponding dissimilarity $D_{ij}$ is defined as:
\begin{equation}
    D_{ij} = f(S_{\text{max}, j}) - f(S_{ij}),
    \label{eq:sim_to_dis}
\end{equation}
where $S_{\text{max}, j} = \max_k(S_{kj})$ is the maximum observed similarity for object $j$, and $f(\cdot)$ is a monotonically increasing function, such as the identity or a logarithmic function. A logarithmic transform, $f(x) = \log(x)$, is appropriate when similarities span several orders of magnitude (like X-fold changes) and relative differences are more meaningful than absolute ones.
\end{definition}
This transformation ensures that the highest similarity corresponds to a dissimilarity of zero (i.e., a distance of zero), and lower similarities correspond to larger dissimilarities.

\subsection{The Physical Model}
Topolow models the set of objects as a system of $m$ particles in $\R^N$. The interactions between particles are governed by laws of motion in physics \citep{newton1687principia}. This approach is conceptually similar to the "spring-embedder" model from force-directed graph drawing \citep{kobourov2012}, but with crucial differences in its objective (see Remark \ref{remark:force}).

\begin{definition}[Potential Energy and Forces in Topolow]
The total potential energy $U(X)$ of a configuration $X = \{\textbf{x}_1, \dots, \textbf{x}_{m}\}$ is the sum of two types of potentials:
\begin{enumerate}
    \item \textbf{Spring Potential:} For each pair $(i,j)$ in the observed data $\mathcal{O}$, a Hookean spring connects particles $i$ and $j$. The spring has a rest length of $D_{ij}$ and a spring constant $k$. The potential energy is:
    \begin{equation}
        U_{s,ij}(x_i, x_j) = \frac{1}{2} k \left( \norm{\textbf{x}_i - \textbf{x}_j} - D_{ij} \right)^2
    \end{equation}
    \item \textbf{Repulsive Potential:} For any pair of particles $(a,b)$ with no specified dissimilarity (i.e., a missing measurement), the particles exert a repulsive force on each other. This is modeled by an inverse-distance potential with a repulsion constant $c$:
    \begin{equation}
        U_{r,ab}(x_a, x_b) = \frac{c}{\norm{\textbf{x}_a - \textbf{x}_b}}
    \end{equation}
\end{enumerate}
The total potential energy of the system is $U(X) = \sum_{(i,j) \in \mathcal{O}} U_{s,ij}(X) + \sum_{(a,b) \notin \mathcal{O}} U_{r,ab}(X)$. 

\end{definition}

The forces derived from these potentials are $\mathbf{F}_{s,ij}$ and $\mathbf{F}_{r,ab}$, respectively. Specifically:
\begin{align}
    \mathbf{F}_{s,ij} &= -k \left( \norm{\textbf{x}_i - \textbf{x}_j} - D_{ij} \right) \frac{\textbf{x}_i - \textbf{x}_j}{\norm{\textbf{x}_i - \textbf{x}_j}}, \\
    \mathbf{F}_{r,ab} &= \frac{c}{\norm{\textbf{x}_a - \textbf{x}_b}^2}\frac{\textbf{x}_a - \textbf{x}_b}{\norm{\textbf{x}_a - \textbf{x}_b}}.
\end{align}
The spring force pulls or pushes particles $i$ and $j$ so their Euclidean distance approaches $D_{ij}$. The repulsive force ensures that particles with no observed relationship do not collapse onto each other, a common failure mode in sparse embeddings. The repulsive forces and potentials are negligible when added to springs' forces and potentials.

\begin{remark}\label{remark:force}[Distinction from Graph Drawing]
A key distinction between Topolow and classical force-directed graph drawing lies in the definition of the spring rest lengths. In methods like \citet{fruchterman1991}, all springs have a uniform ideal length, aiming to space connected nodes evenly. In methods like \citet{kamada1989}, the rest length is proportional to the graph-theoretic shortest path distance. The goal of these methods is an aesthetically pleasing and readable layout of a graph's topology. In contrast, Topolow uses the specific, quantitative dissimilarity values $D_{ij}$ as the rest lengths. Its objective is not merely a layout, but a precise \textbf{metric reconstruction} where the geometric distances in the embedding space quantitatively match the input dissimilarities.
\end{remark}

\subsection{Statistical Foundation and Loss Function}
\label{sec:loss}

Topolow's objective is to minimize the Mean Absolute Error (MAE) between the observed dissimilarities and the embedded Euclidean distances:
\begin{equation}
    \text{MAE}(X) = \frac{1}{|\mathcal{O}|} \sum_{(i,j) \in \mathcal{O}} \lvert D_{ij} - \norm{\textbf{x}_{ij}}\rvert,
\end{equation}
where $\norm{\textbf{x}_{ij}} = \norm{\textbf{x}_i - \textbf{x}_j}$. This choice is not arbitrary; it has a firm statistical basis.
\begin{theorem}
Minimizing the MAE loss function is equivalent to finding the Maximum Likelihood Estimate (MLE) for the configuration $X$ under the assumption that the measurement errors, $\epsilon_{ij} = D_{ij} - \norm{\textbf{x}_{ij}}$, are independent and identically distributed according to a Laplace distribution with mean 0.
\end{theorem}

\begin{proof}
The probability density function (PDF) of a Laplace distribution with location $\mu$ and scale $b$ is $f(\epsilon|\mu,b) = \frac{1}{2b} \exp\left(-\frac{|\epsilon - \mu|}{b}\right)$. Assuming errors are centered at zero ($\mu=0$), the log-likelihood for a set of observations $\mathcal{O}$ given a configuration $X$ is:
\begin{align}
    \log \mathcal{L}(X, b | D) &= \log \prod_{(i,j) \in \mathcal{O}} \frac{1}{2b} \exp\left(-\frac{\lvert\norm{\textbf{x}_{ij}} - D_{ij}\rvert}{b}\right) \\
    &= \sum_{(i,j) \in \mathcal{O}} \left( -\log(2b) - \frac{\lvert\norm{\textbf{x}_{ij}} - D_{ij}\rvert}{b} \right) \\
    &= -|\mathcal{O}|\log(2b) - \frac{1}{b} \sum_{(i,j) \in \mathcal{O}} \lvert\norm{\textbf{x}_{ij}} - D_{ij}\rvert\label{eq:logL}
\end{align}
The MLE for the scale parameter, $b$, is the MAE:
\begin{equation*}
\hat{b} = \frac{1}{|\mathcal{O}|} \sum_{(i,j) \in \mathcal{O}} |\epsilon_{ij} | = \frac{1}{|\mathcal{O}|} \sum_{(i,j) \in \mathcal{O}} \lvert\norm{\textbf{x}_{ij}} - D_{ij}\rvert = \text{MAE}.
\end{equation*}
Plugging this into equation~(\ref{eq:logL}) we get:
\begin{equation}
    \log \mathcal{L}(X| D) = - |\mathcal{O}|\log(2 \cdot \text{MAE}) - |\mathcal{O}|.
\end{equation}
To maximize the log-likelihood with respect to $X$, we must minimize the MAE. 
\end{proof}

The Laplace distribution is more robust to outliers than the Gaussian distribution (which corresponds to an $L_2$ or squared-error loss) due to its heavier tails, making this framework well-suited for noisy experimental data.

\subsubsection{Likelihood for Censored Data}
Real-world data sometimes contain censored measurements (\eg values below a detection limit, $c^L$, recorded as `$<$$c^L$`, or above a saturation limit, , $c^R$, recorded as `$>$$c^R$`). Topolow incorporates these directly into the likelihood function.

Let the error be $\epsilon_{ij} = D_{ij} - \norm{\textbf{x}_{ij}}$. For a right-censored observation $D_{ij} > c^R_{ij}$, the true error is $\epsilon_{ij} > c^R_{ij} - \norm{\textbf{x}_{ij}}$. Its contribution to the likelihood is the survivor function $S(c^R_{ij} - \norm{\textbf{x}_{ij}}) = P(\epsilon_{ij} > c^R_{ij} - \norm{\textbf{x}_{ij}})$. For a left-censored observation $D_{ij} < c^L_{ij}$, the true error is $\epsilon_{ij} < c^L_{ij} - \norm{\textbf{x}_{ij}}$, and its contribution is the CDF $F(c^L_{ij} - \norm{\textbf{x}_{ij}}) = P(\epsilon < c^L_{ij} - \norm{\textbf{x}_{ij}})$.

For a Laplace($0, b$) distribution, the CDF is $F(z) = \frac{1}{2}\exp(z/b)$ for $z < 0$, and $F(z) = 1 - \frac{1}{2}\exp(-z/b)$ for $z \ge 0$. The survivor function is $S(z)=1-F(z)$.

The full log-likelihood for a dataset with exact ($\mathcal{O}_{exact}$), left-censored ($\mathcal{O}_{left}$), and right-censored ($\mathcal{O}_{right}$) observations is:
\begin{equation}
\begin{split}
    \log \mathcal{L} = & \sum_{(i,j) \in \mathcal{O}_{exact}} \log f(D_{ij}-\norm{\textbf{x}_{ij}}) \\
    & + \sum_{(i,j) \in \mathcal{O}_{left}} \log F(c^L_{ij}-\norm{\textbf{x}_{ij}}) \\
    & + \sum_{(i,j) \in \mathcal{O}_{right}} \log S(c^R_{ij} - \norm{\textbf{x}_{ij}})
\end{split}
\end{equation}
where $c^L_{ij}$ and $c^R_{ij}$ are the censoring thresholds.\newline

Implementation of the Topolow algorithm involves two distinct phases: first, estimation of the model's hyperparameters, and second, optimization of the particle coordinates using those parameters. Both phases are grounded in a likelihood framework.

\subsection{Parameter Estimation via Adaptive Monte Carlo}
\label{sec:param_est}
Prior to the sequential optimization to find particle coordinates, Topolow finds the maximum likelihood estimates of the model's parameters $\theta = \{N, k_0, c_0, \alpha\}$ for the input data. At this stage, Topolow's objective is to find the parameter set $\theta$ that maximizes the likelihood of the data, typically evaluated using $k$-fold cross-validation. The log-likelihood derived from the MAE on held-out validation data serves as a principled objective function for this optimization:
\begin{equation}
    \log \mathcal{L}(X_{val} | D) = -|\mathcal{O}_{val}|\log(2 \cdot \text{MAE}_{val}) - |\mathcal{O}_{val}|.
\end{equation}

This search of the parameter space is performed using a two-stage sampling process:
\begin{enumerate}
    \item \textbf{Initial Exploration with Latin Hypercube Sampling (LHS):} To ensure the parameter space is explored broadly, an initial set of parameter combinations is generated using LHS \citep{mckay1979comparison}. LHS is a stratified sampling method that provides more uniform coverage of a multi-dimensional parameter space than simple random sampling.
    \item \textbf{Adaptive Monte Carlo (AMC) Refinement:} After evaluating the initial LHS samples, the algorithm transitions to an adaptive sampling phase \citep{bucher1988}. In each step of AMC, a kernel density estimator (KDE) is used to approximate the likelihood surface based on all previously evaluated samples. New candidate parameter sets are then drawn from this estimated distribution, focusing the search on regions of high likelihood. This adaptive process allows for efficient convergence to the maximum likelihood estimate of the hyperparameters.
\end{enumerate}

A critical parameter in any embedding is the dimension, $N$. Choosing $N$ too small can lead to significant distortion (high stress), while choosing it too large can lead to overfitting and poor generalization. Topolow treats $N$ as a model parameter to be optimized alongside the other parameters, thereby selecting the most statistically appropriate dimension for the data. The final output of this stage is the optimal parameter set $\theta^*$ that will be used to generate the final map.

\subsection{The Gradient-Free Optimization Algorithm}
\label{sec:algo}
With the optimal hyperparameters $\theta^*$ determined, the final map is generated by running the core optimization algorithm on the full dataset.
The algorithm proceeds as follows:
\begin{enumerate}
    \item \textbf{Initialization:}
    \begin{enumerate}
        \item Reorder the dissimilarity matrix to place objects with the highest average dissimilarities at opposing ends of the matrix. This spectral-like ordering provides a more stable starting structure to improve convergence.
        \item Assign initial random coordinates $x_k \in \R^N$ to all particles, across the spectral of the dissimilarity matrix.
        \item Use the optimal parameters $k_0, c_0, \alpha$ found in Section \ref{sec:param_est}.
    \end{enumerate}
    \item \textbf{Iteration:} For each iteration $t=1, 2, \dots$ until convergence:
    \begin{enumerate}
        \item Create a random permutation of all unique pairs of particles $(a,b)$.
        \item For each pair $(a,b)$ in the permutation, update the positions of $x_a$ and $x_b$ based on their single pairwise force, holding all other particles fixed, as derived in Proposition \ref{prop:displacement}.
        \item Update (cool) the parameters: $k_t = k_0(1-\alpha)^t$ and $c_t = c_0(1-\alpha)^t$.
        \item Check for convergence based on the relative change in the MAE.
    \end{enumerate}
\end{enumerate}

\begin{proposition}[Particle Displacement]
\label{prop:displacement}
In an iteration, the displacement of particle $a$ due to its interaction with particle $b$ has magnitude $d_a$ and is directed along the vector connecting the particles:
\begin{align}
    d_{s,a} &= \frac{2k(\norm{\textbf{x}_{ab}} - D_{ab})}{4m_a + k} \quad \text{if } (a,b) \in \mathcal{O} \\
    d_{r,a} &= \frac{c}{2m_a \norm{\textbf{x}_{ab}}^2} \quad \text{if } (a,b) \notin \mathcal{O}.
\end{align}
The term $m_a$ is an effective mass, typically proportional to the number of observed dissimilarities for particle $a$.
\end{proposition}
\begin{proof}
The following derivation is an elaboration on ideas from \citet{arhami2025topolow}. The interaction is modeled as a one-dimensional motion along the line connecting particles $a$ and $b$. We calculate the displacement of particle $a$, denoted $d_{s,a}$, while particle $b$ is held fixed.

For a spring force, the system starts at rest ($v_t=0$) with distance $r_t = \norm{\textbf{x}_a - \textbf{x}_b}$. After a time step $\Delta t=1$, particle $a$ has moved a distance $d_{s,a}$ to a new position, resulting in a new distance $r_{t+1} = r_t - d_{s,a}$ and a final velocity $v_{t+1}$. Under constant acceleration, the displacement is $d_{s,a} = \frac{v_t + v_{t+1}}{2}\Delta t = \frac{v_{t+1}}{2}$, which implies $v_{t+1} = 2d_{s,a}$.

By the work-energy theorem, the change in potential energy of the spring is converted into the kinetic energy of the particle:
\begin{equation}
    U(r_t) - U(r_{t+1}) = \frac{1}{2}m_a v_{t+1}^2.
\end{equation}
Substituting the spring potential energy $U(r) = \frac{1}{2}k(r - D_{ab})^2$:
\begin{equation}
    \frac{1}{2}k(r_t - D_{ab})^2 - \frac{1}{2}k(r_{t+1} - D_{ab})^2 = \frac{1}{2}m_a v_{t+1}^2.
\end{equation}
Now, substitute $r_{t+1} = r_t - d_{s,a}$ and $v_{t+1} = 2d_{s,a}$:
\begin{equation}
    k \left[ (r_t - D_{ab})^2 - (r_t - d_{s,a} - D_{ab})^2 \right] = m_a (2d_{s,a})^2.
\end{equation}
Let $X = r_t - D_{ab}$. The equation becomes:
\begin{align}
    k \left[ X^2 - (X - d_{s,a})^2 \right] &= 4m_a d_{s,a}^2 \\
    k \left[ X^2 - (X^2 - 2Xd_{s,a} + d_{s,a}^2) \right] &= 4m_a d_{s,a}^2 \\
    k (2Xd_{s,a} - d_{s,a}^2) &= 4m_a d_{s,a}^2.
\end{align}
Assuming $d_{s,a} \neq 0$, we can divide by it: 
\begin{align}
    k (2X - d_{s,a}) &= 4m_a d_{s,a} \\
    2kX &= (4m_a + k) d_{s,a}.
\end{align}
Solving for the displacement $d_{s,a}$:
\begin{equation}
    d_{s,a} = \frac{2kX}{4m_a + k} = \frac{2k(r_t - D_{ab})}{4m_a + k}.
\end{equation}
This gives the magnitude of the displacement for particle $a$. The algorithm finds $d_{s,b}$ by replacing $m_a$ with $m_b$.

For the repulsive force, $F_{r,ab} = c / r_t^2$. Assuming this force is constant over a small displacement, the acceleration is $a_{r,a} = F_{r,ab}/m_a$. The displacement is $d_{r,a} = \frac{1}{2}a_{r,a}(\Delta t)^2 = \frac{a_{r,a}}{2} = \frac{c}{2m_a r_t^2}$.
\end{proof}

\subsection{Software Availability}
The Topolow algorithm is implemented in the R package \texttt{topolow}, available from CRAN. The main function \texttt{Euclidify} accepts dissimilarity matrices and returns Euclidean coordinates, with automatic hyperparameter optimization and dimension selection. The package includes comprehensive documentation, examples, and supporting functions for data pre-processing.

\section{Empirical Performance Evaluation on a General Dataset}
\label{sec:empirical_eval}
To rigorously evaluate the performance of Topolow, we designed a series of simulation experiments using synthetic data. This approach allows us to control the ground-truth properties of the data—including its geometric complexity, size, and sparsity—providing a clear and objective benchmark against which to measure performance.

\subsection{Experimental Design and Methodology}
\subsubsection{Data Generation}
We generated a series of synthetic dissimilarity matrices with known challenging properties. The process begins by creating a set of ground-truth coordinates for $m=50$ objects in a high-dimensional space ($\R^{10}$) with a predefined cluster structure. From these coordinates, a foundational Euclidean distance matrix is computed and systematically distorted through a series of non-linear transformations, including range-dependent power transformations and the addition of asymmetric noise. This procedure yields a complete but non-metric dissimilarity matrix, $M$, which serves as the ground truth for our evaluations. Then, we randomly removed 30\% of the pairwise dissimilarities to introduce missing values and used the resulting matrix ($D$) as the input of the algorithms.

To test robustness under different conditions, we conducted two main sets of experiments:
\begin{enumerate}
    \item \textbf{Multi-Sparsity Analysis:} For a fixed number of objects ($m=50$), we introduced missing values by randomly removing 30\%, 60\%, and 90\% of the pairwise dissimilarities to create three variants of $D$.
    \item \textbf{Multi-Scale Analysis:} For a fixed sparsity level (30\% missing), we generated three variants of $D$ with $m=25, 50,$ and $100$ objects.
\end{enumerate}

\subsubsection{Comparison Methods}
We compared the performance of Topolow against two standard MDS implementations, representing different algorithmic philosophies:
\begin{enumerate}
    \item \textbf{Classical MDS:} A deterministic method based on the eigenvalue decomposition of the Gram matrix, as implemented in the R function \texttt{cmdscale} \citep{torgerson1952multidimensional, gower1966some}. Since this method requires a complete dissimilarity matrix, missing values in the training set were imputed using the median of the available dissimilarities.
    \item \textbf{Iterative Metric MDS:} A stochastic optimization method that seeks to minimize a STRESS objective function. We used the implementation in the R package \texttt{smacof} \citep{Leeuw2009}. This method also required median imputation for missing values.
\end{enumerate}
For Topolow, we used the \texttt{Euclidify} function from the R package \texttt{topolow}, with hyperparameters for each experimental condition determined using the adaptive Monte Carlo procedure described in Section  \ref{sec:param_est}. For the stochastic methods (Topolow and Iterative MDS), each experiment was repeated three times to assess the stability and consistency of the solutions.

\subsubsection{Evaluation Metrics}
Performance was assessed by comparing the distances in the resulting Euclidean embedding, $\hat{D}$, to the complete, ground-truth non-metric matrix, $M$.
\begin{itemize}
    \item \textbf{Accuracy:} The primary measure of accuracy was the \textbf{Normalized Stress}, defined as $\sqrt{\sum(M_{ij} - \hat{D}_{ij})^2 / \sum M_{ij}^2}$. Note that unlike $D$, $\hat{D}$ has no missing values. This metric quantifies the overall deviation of the embedded distances from the true dissimilarities. Lower stress indicates a more faithful reconstruction.
    \item \textbf{Distance Preservation:} The linear correlation between true and embedded distances was assessed using the Pearson correlation coefficient ($r$) and the coefficient of determination ($R^2$) from a linear model fit. Visual inspection was performed using Shepard plots, which plot embedded distances against true dissimilarities.
    \item \textbf{Stability:} For stochastic methods, stability was measured by the standard deviation of the Normalized Stress across the replicate runs.
\end{itemize}

\subsection{Data Characteristics}
A critical first step was to quantify the non-metric nature of our synthetic data. By computing the Gram matrix and its eigenvalues, we can measure the deviation from Euclidean geometry. The \textbf{non-metric Deviation Score}, calculated as the ratio of the sum of the absolute values of negative eigenvalues to the sum of positive eigenvalues \citep{gower1985}, provides a standardized measure of this distortion. For our primary test case ($m=50$, 30\% missing), the deviation score was 0.52, with the negative eigenvalues accounting for over a third of the total variance. This confirms the dataset is non-metric, presenting a significant challenge for any embedding algorithm.

\subsection{Comparative Performance Analysis}

\subsubsection{Accuracy and Distance Reconstruction}
In the main test case ($m=50$, 30\% sparsity), Topolow demonstrated substantially higher accuracy in reconstructing the ground-truth dissimilarities compared to both Classical and Iterative MDS. As shown in Table \ref{tab:main_comparison}, Topolow achieved a mean Normalized Stress of $0.190 \pm 0.0003$, an improvement over Classical MDS (0.602) and Iterative MDS (0.552).

\begin{table}[h!]
\centering
\caption{Performance comparison on a non-metric dataset of 50 objects with 30\% missing values. Values for stochastic methods are mean $\pm$ s.d. over 50 runs. Lower stress is better; higher $R^2$ is better.}
\label{tab:main_comparison}
\begin{tabular}{lcc}
\toprule
\textbf{Method} & \textbf{Normalized Stress} & \textbf{Distance Correlation ($R^2$)} \\
\midrule
\textbf{Topolow} & \textbf{0.190 $\pm$ 0.0003} & \textbf{0.804} \\
Classical MDS & 0.602 & 0.705 \\
Iterative MDS & 0.552 $\pm$ 0.0000 & 0.619 \\
\bottomrule
\end{tabular}
\end{table}

\begin{figure}[h!]
    \centering
\includegraphics[width=\textwidth]{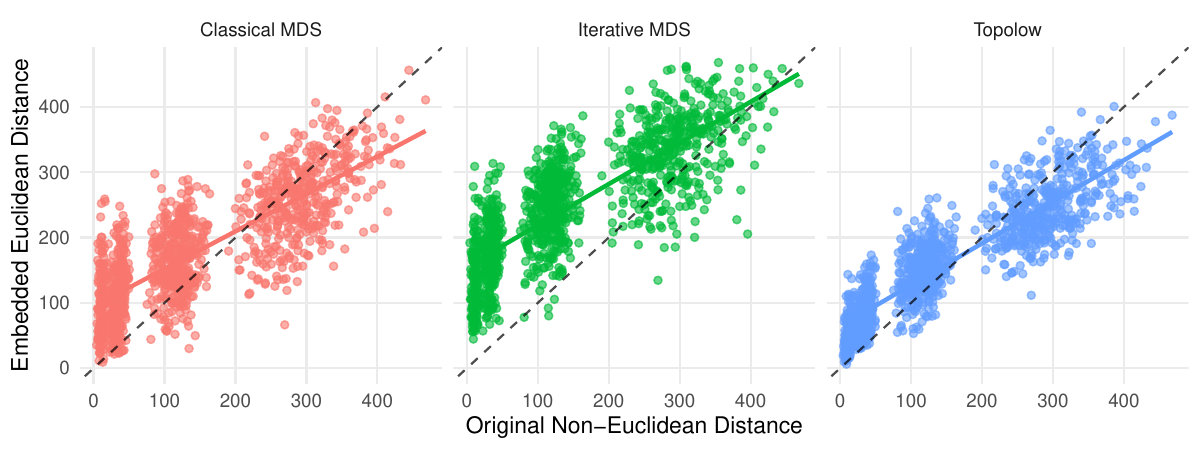}
    \caption{Shepard plots comparing embedded Euclidean distances to original non-metric dissimilarities for the best run of each method ($m=50$, 30\% missing). The identity line (dashed) represents perfect reconstruction. Topolow's embedding (right) shows a strong linear relationship and low distortion. Classical MDS (left) and Iterative MDS (center) show substantially more scatter and deviation from the ideal.}
    \label{fig:shepard_main}
\end{figure}

The superior performance of Topolow is further illustrated by the Shepard plots in Figure \ref{fig:shepard_main}. The points for Topolow show a much tighter, more linear relationship with the identity line ($y=x$) compared to the MDS methods. This improved fidelity stems from Topolow's optimization process, which leverages the entire network of known dissimilarities simultaneously. Rather than relying on pre-imputed values, it finds a configuration that minimizes the stress across all known spring-like constraints, resulting in a more coherent geometric structure.

\subsubsection{Robustness to Data Sparsity}
\begin{figure}[h!]
    \centering
\includegraphics[width=0.6\textwidth]{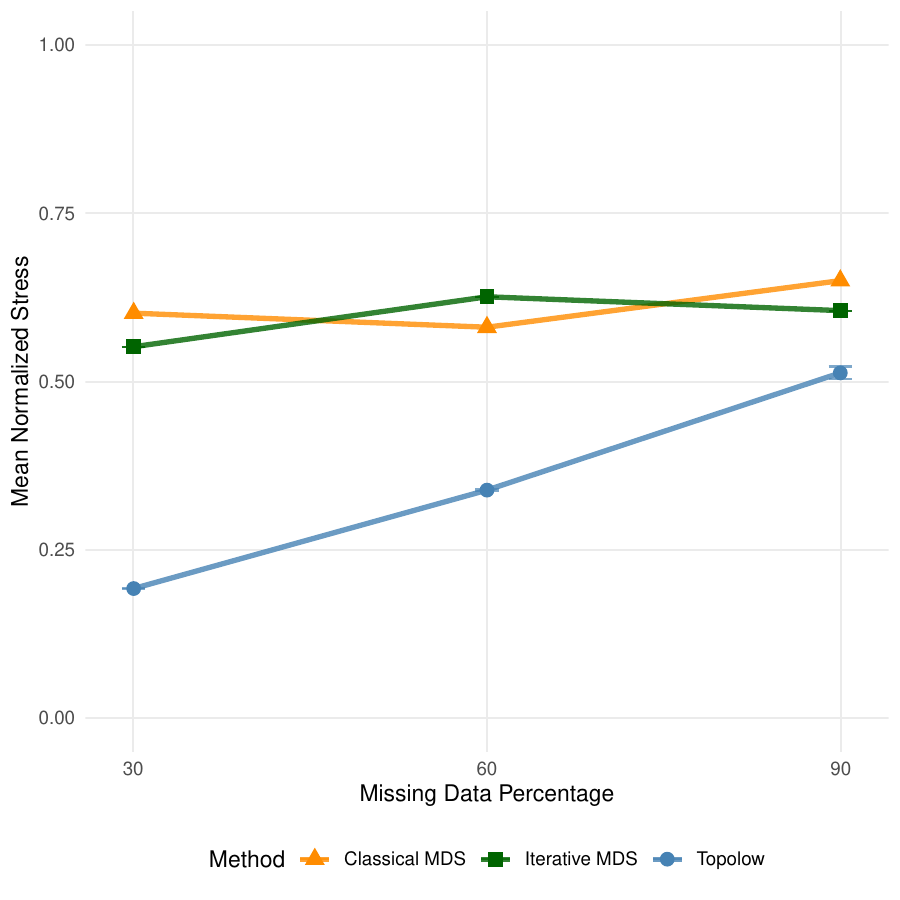}
    \caption{Performance trends across increasing data sparsity for a 50-object dataset. Points show the mean Normalized Stress over 50 runs (error bars show $\pm$ s.d.). Topolow consistently achieves lower stress (better accuracy) than imputation-based MDS methods, and its performance advantage is maintained even at extreme sparsity levels.}
    \label{fig:sparsity_trends}
\end{figure}

To evaluate robustness to missing information, we compared the methods on datasets with 30\%, 60\%, and 90\% sparsity. As summarized in Figure \ref{fig:sparsity_trends}, Topolow consistently and substantially outperformed both MDS methods across all sparsity levels.

At 90\% sparsity—an extreme condition where only 10\% of dissimilarities are known—Topolow achieved a mean Normalized Stress of $0.510$. This result is particularly striking as it is better than the performance of both MDS methods at the much lower 30\% sparsity level. This demonstrates the power of Topolow's approach: the interconnected system of springs enforces a global geometric consistency that allows for robust inference of the overall structure, even when the vast majority of direct measurements are absent. This is fundamentally more effective than attempting to embed a matrix where 90\% of the values have been imputed.

\subsubsection{Scalability with Dataset Size}
\begin{figure}[h!]
    \centering
\includegraphics[width=0.6\textwidth]{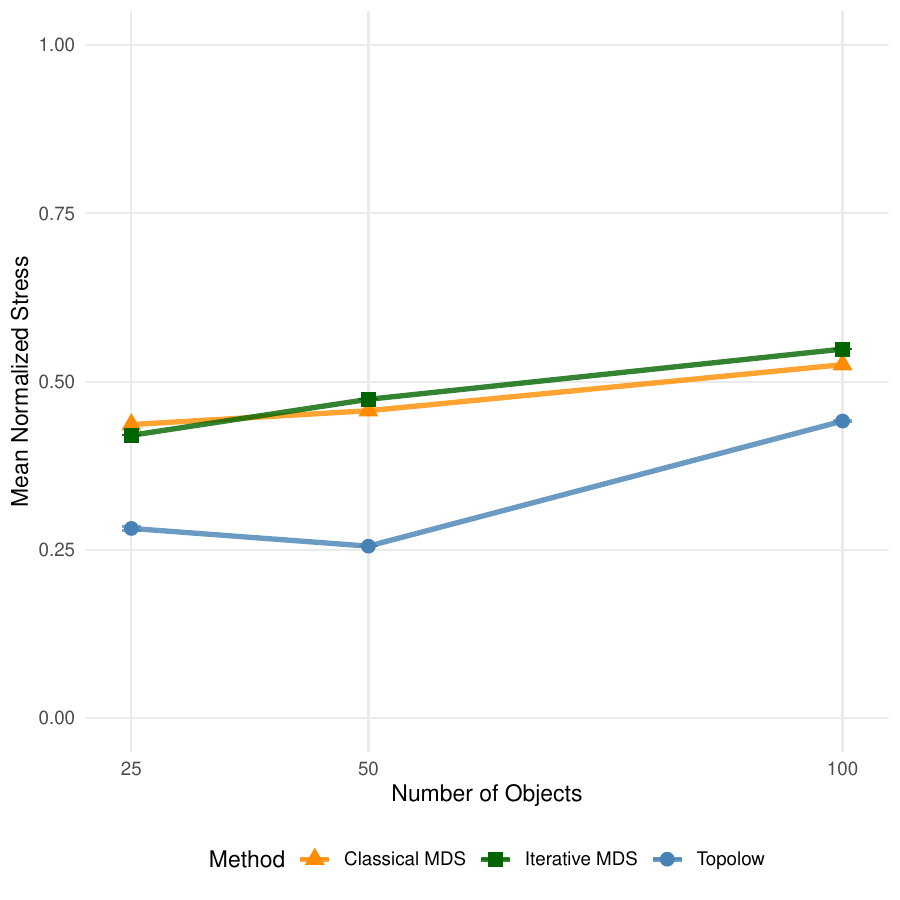}
    \caption{Performance trends across increasing dataset size (25, 50, and 100 objects) with 30\% data sparsity. Points show the mean Normalized Stress over 50 runs (error bars show $\pm$ s.d.). Topolow maintains its superior performance across all tested scales.}
    \label{fig:scale_trends}
\end{figure}

We next assessed performance on datasets of 25, 50, and 100 objects, with sparsity held constant at 30\%. As shown in Figure \ref{fig:scale_trends}, Topolow was the best-performing method at every scale tested. For the largest dataset ($m=100$), Topolow achieved a mean Normalized Stress of 0.441, compared to 0.554 for Classical MDS and 0.567 for Iterative MDS. This demonstrates that the performance advantages of Topolow are not limited to small datasets and that the algorithm scales effectively to larger problems.

\subsection{Stability of Stochastic Methods}
A crucial aspect of any stochastic optimization algorithm is the stability of its solutions. Across all experiments, Topolow produced highly consistent embeddings. The standard deviation of the Normalized Stress over replicate runs was consistently low, on the order of $10^{-3}$ to $10^{-4}$ for lower sparsity levels (see Table \ref{tab:main_comparison} and error bars in Figures \ref{fig:sparsity_trends} and \ref{fig:scale_trends}). Iterative MDS, which is also  a stochastic method, produced identical results in our experiments (s.d. = 0), due to its optimal initialization (init = "torgerson" in smacofSym function) and convergence to the same local minimum. The low variance in Topolow's results suggests that its gradient-free, pairwise optimization scheme, combined with the parameter cooling schedule, robustly navigates the complex energy landscape to find a stable, high-quality solution.

\subsection{Discussion of Results}
The empirical results highlight the practical advantages of Topolow's  design. The consistent outperformance (Table~\ref{tab:main_comparison}), especially in the face of non-metricity and high sparsity, provides strong evidence for the efficacy of its core mechanics.

The robustness of the embeddings (Table~\ref{tab:main_comparison}) to the synthetic, asymmetric noise in the data validates the statistical foundation of the algorithm. The choice of an MAE loss function (equivalent to a Laplace error model) is shown to be not just a theoretical advantage but a practical one, enabling Topolow to reconstruct the underlying geometry more accurately than the $L_2$-based MDS methods when errors are not well-behaved. The combination of this robust statistical model with a holistic, imputation-free optimization framework explains the significant performance gains observed across all tested conditions.

Furthermore, the failure of imputation, as a strategy for handling sparse and non-metric data, has been demonstrated (Table \ref{tab:main_comparison} and Figure \ref{fig:sparsity_trends}). Both MDS methods, which depend on imputation, produced embeddings with substantially higher stress. This suggests that filling in missing values with a statistic, inferred from the rest of the data, corrupts the geometric information encoded in the existing data, leading the optimization astray. In contrast, Topolow's ability to work directly with the sparse matrix, using the system of known dissimilarities to infer the global structure, is clearly a superior approach.

\section{Mathematical Properties and Discussion}

\subsection{The Input Dissimilarity Space}
A key property of the Topolow framework is its generality with respect to the input data.
\begin{proposition}
The Topolow algorithm does not require the input dissimilarity matrix $D$ to be a metric. The dissimilarities $D_{ij}$ are only required to be non-negative real numbers. They need not satisfy:
\begin{itemize}
    \item Symmetry: $D_{ij} = D_{ji}$
    \item The Triangle Inequality: $D_{ik} \le D_{ij} + D_{jk}$
\end{itemize}
\end{proposition}
\begin{proof}
The algorithm's mechanics treat each $D_{ij}$ as an independent target for the spring rest length between particles $i$ and $j$. If $D_{ij} \neq D_{ji}$, the algorithm will simply use the value corresponding to the specific pair $(i,j)$ being considered in a given step. Violations of the triangle inequality are resolved by the system of forces; if $D_{ik} > D_{ij} + D_{jk}$, the springs for $(i,j)$ and $(j,k)$ will pull those particles closer, while the spring for $(i,k)$ pulls them apart. The final configuration $X$ represents a minimum-energy compromise that best fits all constraints simultaneously, effectively finding a Euclidean projection of the non-metric input.
\end{proof}
This makes Topolow applicable to a wide range of data, including psychological ratings, geographic cartography, phenotypic dissimilarities, chemical bindings or cross-reactivity strengths, or other measurements where metric properties are not guaranteed. The algorithm finds the best-fit Euclidean metric representation of this general dissimilarity space. This is a critical feature, as it transforms potentially unusable dissimilarity data into a valid coordinate system where the full suite of standard statistical and machine learning tools, from PCA to k-means clustering, can be validly applied.

\subsection{Robustness and Convergence}
The robustness of the Topolow algorithm stems from several interacting mechanisms. First, the stochastic, pairwise nature of the optimization allows the system to effectively explore the solution space. By updating positions based on individual two-body interactions in a random order, the algorithm is not committed to a single, potentially misleading, global gradient. This allows it to navigate a complex, non-convex energy landscape and avoid getting trapped in shallow local minima.

Second, the system of interconnected springs provides a mechanism for error dampening. A single particle is subject to forces from all of its measured neighbors. If one measurement is an outlier, the corresponding spring will exert a strong, incorrect force. However, this force is counteracted by the collective forces from all other connected springs, which pull the particle towards a consensus position. This network-based averaging organically distributes and mitigates the impact of individual noisy or biased measurements, a property demonstrated in \citet{arhami2025topolow} where Topolow's mapping error was consistently lower than the noise and bias intentionally introduced into the input data.

Third, the use of an MAE loss function, which corresponds to a Laplace error model (Theorem 2.2), makes the optimization inherently robust to outliers. Unlike a squared-error ($L_2$) loss, which heavily penalizes large deviations, the $L_1$ loss gives linear weight to errors, preventing single, grossly inaccurate measurements from dominating the optimization process.

Finally, these mechanisms are guided by a parameter cooling schedule, which facilitates a transition from global exploration to local fine-tuning. In the early stages, large force constants allow for significant adjustments, enabling the system to explore the configuration space $X$ broadly. As the parameters cool, the adjustments become smaller, allowing the system to settle into a stable, low-energy minimum. This combination of stochastic exploration, network-based error dampening, a robust loss function, and an annealing-like cooling schedule ensures a robust convergence to a stable, low-error solution.

\section{Conclusion}
We have presented the mathematics and evaluations of Topolow, an algorithm for Euclidean embedding of potentially sparse, noisy, and non-metric dissimilarity data. By adapting concepts from force-directed graph drawing for the purpose of quantitative metric reconstruction and employing a gradient-free, sequential pairwise optimization, Topolow overcomes the primary limitations of MDS methods. Its foundation in a Laplace error model provides robustness to outliers and censored data, and its integrated, likelihood-based approach to dimensionality selection removes the need for ad-hoc choices. Our empirical evaluation on severely non-metric synthetic data demonstrates that Topolow consistently produces more accurate and stable embeddings than standard MDS methods, particularly as data sparsity and problem size increase.
This theoretical robustness translates to practical advantages, as shown in its original application \citep{arhami2025topolow}, where Topolow achieved comparable accuracy to MDS for well-structured data of antigenic dissimilarities of influenza viruses but demonstrated a 56\% and 41\% improvement for more complex dengue and HIV data, respectively. Furthermore, it exhibited orders of magnitude greater run-to-run stability and, unlike MDS, consistently produced a complete embedding of all viruses, a critical feature for sparse datasets. The algorithm's ability to operate on general, non-metric dissimilarity data is one of its most critical features, providing a necessary bridge between raw, real-world measurements and the vast suite of analytical tools that require metric inputs. This makes Topolow a powerful and widely applicable tool for data analysis, visualization, and dimensionality reduction.

\bibliographystyle{apalike}
\bibliography{references} 

\begin{thebibliography}{}

\bibitem[Arhami and Rohani, 2025]{arhami2025topolow}
Arhami, O. and Rohani, P. (2025).
\newblock Topolow: a mapping algorithm for antigenic cross-reactivity and binding affinity assays.
\newblock {\em Bioinformatics}, 41(7):btaf372.

\bibitem[Borg et~al., 2018]{borg2018applied}
Borg, I., Groenen, P.~J., and Mair, P. (2018).
\newblock {\em Applied multidimensional scaling and unfolding}.
\newblock Springer.

\bibitem[Bravo, 2002]{bravo2002}
Bravo, M. (2002).
\newblock {\em A simulation to evaluate the ability of nonmetric multidimensional scaling to recover the underlying structure of data under conditions of error, method of selection, and percent of missing pairs}.
\newblock {Ph.D.} thesis, The University of Texas at Austin.

\bibitem[Bucher, 1988]{bucher1988}
Bucher, C.~G. (1988).
\newblock Adaptive sampling—an iterative fast monte carlo procedure.
\newblock {\em Structural Safety}, 5(2):119–126.

\bibitem[Burago et~al., 2001]{burago2001course}
Burago, D., Burago, Y., and Ivanov, S. (2001).
\newblock {\em A course in metric geometry}, volume~33.
\newblock American Mathematical Society, Providence.

\bibitem[Cooper, 1983]{cooper1983review}
Cooper, L.~G. (1983).
\newblock A review of multidimensional scaling in marketing research.
\newblock {\em Applied Psychological Measurement}, 7(4):427--450.

\bibitem[Duin, 2005]{duin2005dissimilarity}
Duin, R.~P. (2005).
\newblock {\em The dissimilarity representation for pattern recognition: foundations and applications}, volume~64.
\newblock World Scientific.

\bibitem[Fruchterman and Reingold, 1991]{fruchterman1991}
Fruchterman, T. and Reingold, E. (1991).
\newblock Graph drawing by force-directed placement.
\newblock {\em Software: Practice and Experience}, 21(11):1129--1164.

\bibitem[Gower, 1966]{gower1966some}
Gower, J.~C. (1966).
\newblock Some distance properties of latent root and vector methods used in multivariate analysis.
\newblock {\em Biometrika}, 53(3-4):325--338.

\bibitem[Gower, 1985]{gower1985}
Gower, J.~C. (1985).
\newblock Properties of euclidean and non-euclidean distance matrices.
\newblock {\em Linear Algebra and Its Applications}, 67:81--97.

\bibitem[Holm and Sander, 1996]{holm1996mapping}
Holm, L. and Sander, C. (1996).
\newblock Mapping the protein universe.
\newblock {\em Science}, 273(5275):595--602.

\bibitem[Kamada and Kawai, 1989]{kamada1989}
Kamada, T. and Kawai, S. (1989).
\newblock An algorithm for drawing general undirected graphs.
\newblock {\em Information Processing Letters}, 31(1):7--15.

\bibitem[Khayyam, 1936]{khayyam1936difficulties}
Khayyam, O. (1936).
\newblock {\em Discussion of Difficulties of Euclid}.
\newblock Teheran.
\newblock Edited by T. Erani.

\bibitem[Kobourov, 2012]{kobourov2012}
Kobourov, S. (2012).
\newblock Spring embedders and force directed graph drawing algorithms.
\newblock {\em arXiv preprint arXiv:1201.3011}.

\bibitem[Kruskal, 1964]{kruskal1964multidimensional}
Kruskal, J.~B. (1964).
\newblock Multidimensional scaling by optimizing goodness of fit to a nonmetric hypothesis.
\newblock {\em Psychometrika}, 29(1):1--27.

\bibitem[Leeuw and Mair, 2009]{Leeuw2009}
Leeuw, J.~d. and Mair, P. (2009).
\newblock Multidimensional scaling using majorization: Smacof in r.
\newblock {\em Journal of Statistical Software}, 31(3):1–30.

\bibitem[McKay et~al., 1979]{mckay1979comparison}
McKay, M.~D., Beckman, R.~J., and Conover, W.~J. (1979).
\newblock A comparison of three methods for selecting values of input variables in the analysis of output from a computer code.
\newblock {\em Technometrics}, 21(2):239--245.

\bibitem[Newton, 1687]{newton1687principia}
Newton, I. (1687).
\newblock {\em Philosophi Naturalis Principia Mathematica}.
\newblock Josephi Streater.

\bibitem[Struik, 1958]{struik1958omar}
Struik, D.~J. (1958).
\newblock Omar khayyam, mathematician.
\newblock {\em The Mathematics Teacher}, 51(4):280--285.
\newblock http://www.jstor.org/stable/27955652.

\bibitem[Torgerson, 1952]{torgerson1952multidimensional}
Torgerson, W.~S. (1952).
\newblock Multidimensional scaling: I. theory and method.
\newblock {\em Psychometrika}, 17(4):401--419.

\end{thebibliography}

\end{document}